\documentclass[11pt]{article}
\usepackage[utf8]{inputenc}
\usepackage[left=2.5cm, right=2.5cm, top=2.5cm, bottom=2.5cm]{geometry}
\usepackage{setspace}
\title{Assessments in Education\thanks{Contact: University of Bristol, School of Economics. Address: Priory Road Complex, Priory Road, Bristol, BS8 1TU, United Kingdom. E-mail: h.h.sievertsen@bristol.ac.uk. I thank Simon Burgess and Stefania Simion for helpful comments. The usual disclaimer applies.  }}
\author{Hans Henrik Sievertsen (University of Bristol, VIVE, and IZA)}
\date{July 2022 \\ draft prepared for the Oxford Research Encyclopedia of Economics and Finance}
\usepackage{natbib}
\usepackage{palatino}
\usepackage{amsmath,graphicx}

\begin{document}

\maketitle
\section{Summary}
Assessments such as standardized tests and teacher evaluations of students' classroom participation are central elements of most educational systems. Assessments inform the  student, parent, teacher, and school about the student learning progress. Individuals  use the information to adjust their study efforts and to make guide their course choice. Schools and teachers use the information to evaluate effectiveness and inputs. Assessments are also used to sort students into tracks,  educational programmes, and on the labor market. Policymakers use assessments to reward or penalise schools and parents use assessment results to select schools. Consequently, assessments incentivize  the individual, the teacher, and the school to do well. 

Because assessments play an important role in individuals' educational careers, either through the information or the incentive channel, they are also  important  for efficiency, equity, and well-being. The information channel is important for ensuring the most efficient human capital investments: students learn about the returns and costs of effort investments and about their abilities and comparative advantages. However, because students are sorted into educational programs and on the labor market based on assessment results, students optimal educational investment might not equal their optimal human capital investment because of the signaling value.  Biases in assessments and heterogeneity in access to assessments are sources of inequality in education according to gender, origin, and socioeconomic background. These sources have long-running implications for equality and opportunity. Finally, because assessment results also carry important consequences for individuals' educational opportunities and on the labor market, they are a source of stress and reduced well-being. 

This chapter shows that evidence verifies  many of these conjectures, but also reveals  important unresolved questions. For example the research on the effect of assessments on well-being is particularly scarce, that while several studies document biases in teacher evaluations, we know considerably less about how to reduce and avoid these biases.

Keywords: education, assessments, human capital, inequality

\section{Introduction}
\begin{figure}[ht!]
    \centering
    \includegraphics[width=1\linewidth]{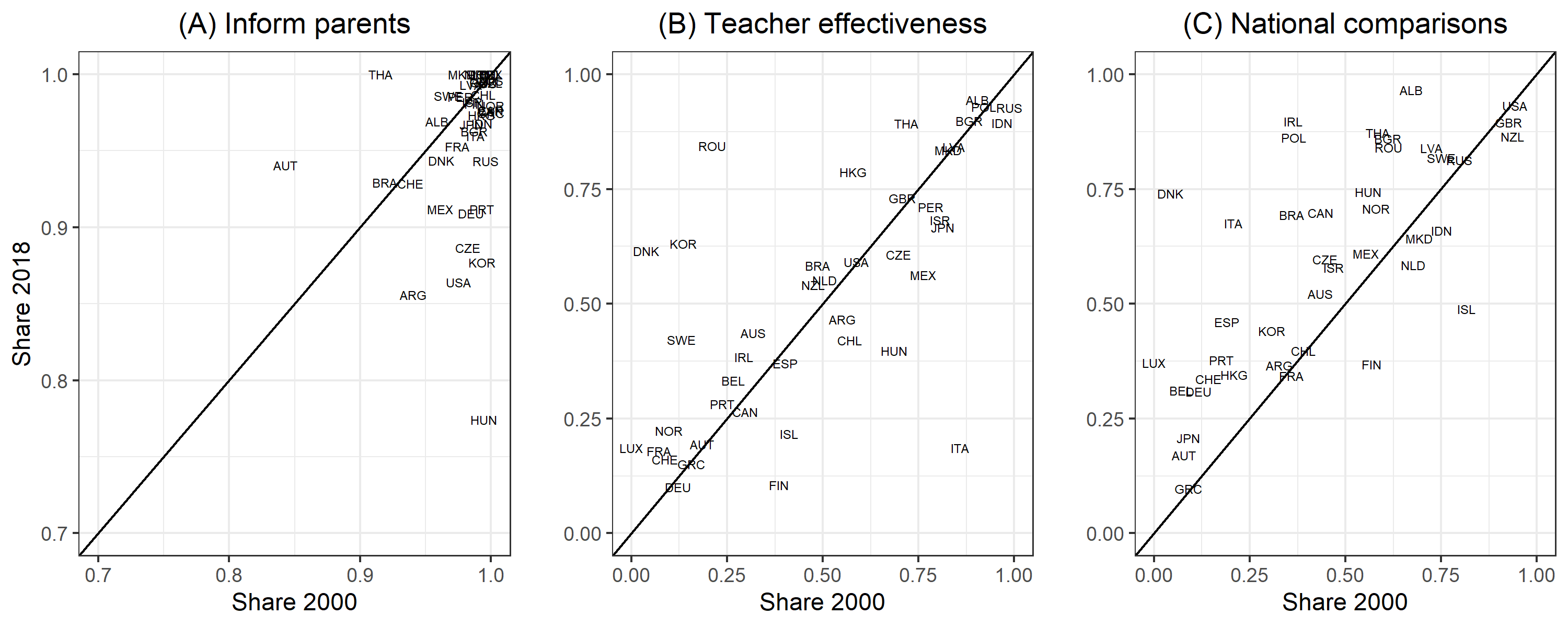}
    \caption{Share of schools using assessments for 15-year-old students to (A) inform parents about their child's progress, (B) make judgments about teacher effectiveness, and (C) compare school to district or national performance. Data sources: PISA 2000 and PISA 2018.}
    \label{fig:fig1}
\end{figure}

Assessments are a central part of most educational systems. As Figure 1(A) illustrates, in both 2000 and 2018, more than 70 percent of schools across 42 countries used assessments to inform parents about their child's progress. Nevertheless the use of assessments varies widely across time and space. For instance, as depicted in Figure 1(B), the frequency at which assessments were used to evaluate teacher effectiveness ranged broadly across countries from from less than ten percent to more than 90 percent across countries. And Figure \ref{fig:fig1}(C) shows that most countries increased the use of assessments for national comparisons over the last two decades. The great degree of heterogeneity in the use of assessments and the changes over time indicate that no consensus has been reached yet on the best assessment system.  The effects that assessments have on individuals and schools must first be determined before an optimal assessment system can be identified.  While education researchers and sociologists   have studied assessments for many decades, economists have recently also demonstrated  interest. This chapter aims at summarising the recent developments in research on  assessments  from the economist's viewpoint. 

Why should economists care about assessments in education? The short answer is that assessments affect efficiency,  equity, and well-being.  Assessments are central components  in school choice and accountability systems that are commonly used to maintain and improve school effectiveness \citep{burgess2021school}.  They affect drop-out decisions and course choices \citep{stinebrickner2012learning,bar2009grade,ahn2019equilibrium} and   are used to sort individuals into educational programs. Hence, providing access to (fair) assessments can therefore reduce inequality in access to education  \citep{hyman2017act}. However, popular assessment types suffer from biases based on gender, origin, and socioeconomic background; furthermore assessments cause increased stress and affect mental health negatively \citep{bach2020understanding,heissel2021testing}. Understanding how assessments affect individuals is, therefore, important to understanding 
human capital development, equity, and well-being. 

The longer explanation for why economists should care about assessments in education relates to the core economic principles underlying many components of assessments and to economists' tools being well-suited for studying assessments. For example, how assessments affect students, teachers, schools, and parents can be studied in light of a principal-agent model \citep{bergbauer2021testing}, where parents act as principals and  schools and teachers act as agents. However, the school and teachers also act as principals where the agent is the student. The teacher cannot perfectly observe the effort of the student, and the parent cannot perfectly observe the effort of the teacher or school. The solution is a "contract" based on assessments, the results of which may affect the school's funding directly in an accountability system or indirectly through school competition and user choice. For students, the "contract" means that the assessment results are important for their future educational track and labor market success, which gives them incentive to study but also has consequences for their mental health and well-being.

The trade-off between the benefits of assessments for school effectiveness and equity and the potential costs in terms of mental health and well-being illustrates that no easy solution exists, and that detailed research across the social sciences that includes evidence from educational sciences is needed. This demand for research is even clearer considering the increased policy attention being placed on assessments. In the 2018 World Development Report, increased use of assessments is included as one of three key strategies to improve global human capital development \citep{wb2018}. However, the increased use of assessments has not been without controversy; for example, critics say it is harmful to learning  \citep{schleicher2014oecd}. The worry that assessments, and  especially standardized testing, may distort students' incentives is not new. Indeed, in the 1880s, Harvard University replaced a finer grading system with the six letters, A-F, that are still used today to reduce the discretionary effect of grades on student motivation \citep{grant2013grades}:
\begin{quote}
   \emph{ "The Faculty last year did away with the minute percentage system of marking, and
substituted a classification of the students in each course of study in five groups,
the lowest of which includes those who have failed in the course. It is hoped
that this grouping system will afford sufficient criteria for the judicious award of
scholarships, honorable mention, and the grade of the Bachelor’s degree, while
it diminishes the competition for marks and the importance attached by students
to College rank in comparison with the remoter objects of faithful work." (Annual
Report of the President of Harvard 1885)}\footnote{This quote originally appeared in  \citet{smallwood1969historical}, but here it is taken from  \citep{grant2013grades}.}
\end{quote}

The controversy over the use of assessments calls for more research on the consequences of assessments and the circumstances under which  assessments might be more harmful than beneficial. And the toolbox of social scientists is excellently suited for that. The chapter is divided into two main parts. The first section includes information on the skills and knowledge measured by assessments and issues about measurement, such as biases, exogenous factors, manipulation, and cheating. The second section presents an overview of the effects that assessments have on individuals, including the importance for learning, well-being, and long-term outcomes.

Several aspects of assessments must be clarified before discussing this topic. First, in the context of this paper, all attempts to evaluate student learning and understanding are considered assessments.  These attempts include standardized tests, oral examinations, teacher assessments of classroom participation, essays, and many other forms of evaluations.  Unless otherwise stated, a standardized test refers to a centrally designed and evaluated assessment and grades are teacher evaluations of student learning.  Second, internal assessments may be teacher assessments of classroom participation or internal tests designed and conducted by the teacher, while external  assessments are often centrally designed and evaluated to be comparable across classrooms, schools, and  even countries.  Third, assessments can be blind or non-blind. In blind assessments, the assessor has no information  about the student, but the assessor either directly observes  the student, knows the student,  or knows characteristics of the student  in non-blind assessments. Fourth, assessments may be internally or externally evaluated. An assessment might be blind, but still be evaluated internally within the same school, in contrast to external assessments that are assessed by staff at other schools or by central institutions.  Fifth, assessments can have varying degrees of  importance for  students and  schools, also called stakes. The outcomes of low-stakes assessments have no direct consequences for students or schools; however, the outcomes of high-stakes assessments may determine whether a student can enroll in the educational program of choice or may affect school funding. 

This chapter is written from an economist’s perspective. As such, the literature covered is primarily based on contributions from economists. While research from the social sciences discipline is also addressed, this chapter does not do justice to the contributions that scholars in the social sciences and other disciplines have made to this topic. As another example, educational research has studied assessment for decades, long before economists cared about the topic. However, studies from this area of research are not covered in this overview either, not because they are not important but to focus the content on the economist viewpoint. Moreover, insights from research that focuses specifically on the measurement, interpretation, and use of assessments and psychometrics also are not covered detail. Nevertheless, several insights and recent contributions by economists are built on the shoulders of insights from these disciplines.

\section{Measurement}
The core intention of an assessment is typically to evaluate students' learning and understanding  in order to give accurate feedback to the student, parents, teachers, the school, and possibly the accountability system. However, the results of an assessment  may also depend on  factors other than student learning and understanding,  such as the amount of effort the student invested in the assessments, the leniency of the assessor, biases, manipulation and cheating, and other external factors like noise and air quality.  The following subsections identify various assessments and explain the skills and knowledge they evaluate, supported by evidence from the literature.

\subsection{What skills assessments measure}
In the context of education, the objective of an assessment is typically to evaluate the learning of the student in contrast to  measuring an innate and constant level of  skill.  Moreover, at least traditionally, the focus of assessments in education has been measuring cognitive skills in contrast to non-cognitive skills, such as grit, locus of control, motivation, interpersonal and social skills. \citet{borghans2016grades} used four datasets representing individuals in the United States, the United Kingdom, and the Netherlands to shed light on   what assessments  measure. The four datasets included results from Intelligent Quotient (IQ) measures, standardized tests, school grades, and personality tests. \citet{borghans2016grades} examined the correlation between the four measures within each sample and showed  that the lowest correlations were found between IQ and grades and IQ and personality. The strongest correlations were found between IQ and results from  standardized test and between grades and standardized tests. The correlation between grades and personality was also reasonably high. The authors concluded that these concepts measure different  aspects of performance: While grades seem to capture elements of both of cognitive skills (strong correlation with standardized tests) and non-cognitive skills (strong correlation with personality), the standardized tests seem to capture mostly cognitive skills (weaker correlation with personality). \citet{borghans2016grades} also studied how assessments predict future assessment outcomes; the findings are in line with the takeaways from the correlations. Finally, the authors also study how the various assessments predict later life outcomes: Grades and standardized test scores predicted later life outcomes, such as wages, health (BMI), and life satisfaction, better than IQ, possibly because grades  also capture personality \citep{borghans2016grades}. The conclusion that school grades capture both cognitive and non-cognitive skills while standardized tests mainly capture the former agree with the conclusions in reviews in the psychological and educational and pedagogical literature (see  \citet{almlund2011personality} for a detailed review).  The finding that grades and standardized tests capture slightly different aspects may explain why many educational systems still apply both types of assessments. 

Recent studies have identified what  test scores do not capture, or more specifically, how teachers affect student behaviors and non-cognitive skills. For example \citet{jackson2018test}, quantified measures of behavior such as absences, suspensions, and grade repetition, and found that they were stronger predictors of later life outcomes than test scores. Importantly, \citet{jackson2018test} showed that teacher value added in terms of test scores were only  weakly correlated with teacher value added in behaviors. In other words, teachers who are good at leading their students to improved test scores are not necessarily good at helping them to improve their non-cognitive skills. 

\subsection{Effort}
Assessments will only give a good indication of students' understanding if the students actually attempted to show their level of comprehension of the material. However, when a test is not important to students, they may choose to invest little effort into demonstrating their understanding. This concern is especially relevant for low-stakes tests that carry little importance for the students themselves and for which the invested effort depends on the students' intrinsic motivation. This has led to the criticism of using low-stakes assessment as a tool to assess  learning, especially across contexts where  intrinsic motivation levels may might vary across the groups compared. One example is the Programme for International Student Assessment (PISA) that is administered in more than 60 countries by the Organisation for Economic Co-operation and Development (OECD) to measure mathematics, reading, and science skills among 15-year-old students. Using the PISA scores to compare the learning across countries relies on assumptions of similar investments in test effort and, thus, similar levels of intrinsic motivation across these countries. \citet{gneezy2019measuring} studied this by randomly adding extrinsic financial incentives to do well to  students conducting a similar test in the United States and in China. The authors reported that the test scores for the Chinese students did not improve as a result of the added extrinsic incentives, but the financial rewards caused an improvement in the American students' performance. This pattern aligns with the hypothesis that the Chinese students had already maximized their effort because of their high level of intrinsic motivation to do well on the low-stakes test. In other words, adding another incentive for these students did not improve the performance because they already were at the limit. The American students, on the other hand, did not exert their maximum test effort because they had lower levels of intrinsic motivation, so they had room to improve their score in response to the incentive. Thus,  different levels of intrinsic motivation across countries may explain some of the variation in PISA results across countries. 

Another approach to studying the role of effort in assessment is to link students' responses to other behavior in the same setting. For example,  the PISA assessment is linked to an extensive survey. \citet{fuchsman2020testing} investigated inconsistent answers to the survey questions. For instance, a student may indicate liking school in response to one survey question but may indicate not liking school in response to a subsequent question that is worded slightly differently; this outcomes would suggest a lack of effort and lack of attention to the survey on the student's part.  Using this measure of disengagement with the PISA survey as well as a measure based on item non-responses to the survey, \citet{fuchsman2020testing} concluded that variation in effort can explain between  32 and 38 percent of the variation in PISA test scores across countries.

\subsection{Grading standards}
Standardized tests are fairly straightforward to ensure consistent standards across time and contexts. These tests may even be multiple choice assessments with computer-based evaluations. However, as described previously, these tests do not measure all aspects of student learning and assessments involving human judgment, such as teacher grades, play an important role in many contexts. Due to the subjectivity and human aspect of these assessments, the grade a student receives for a given performance may vary across assessors, time, and contexts. Related issues, such as grade inflation, grade compression, and general variation in grading standards, are discussed in the following subsection.

Aggregate statistics provide clear evidence of so-called grade inflation in many countries and settings. This is typically seen as an increase in grade point averages or an increasing share of a cohort receiving a specific degree classification. For example, in the United Kingdom, the share of students receiving the highest degree classification (a "first" or "distinction")  increased from 15.7 percent in 2011 to 37.9 percent in 2021 \citep{ofs2022}. In the United States, the share of high school students with top grades increased from 39 percent in 1998 to 47 percent in 2016 \citep{buckley2018measuring}. While these changes may reflect changes in student composition,  evidence suggests that this is the result of more lenient marking. For instance the standardized SAT scores of high school students in the United States did not improve similarly to the increase in grades from 1996 to 2016 \citep{buckley2018measuring}. Using data on grades from 1982 to 2001 at Clemson University, \citet{hernandez2016measuring} showed that about one-quarter of the grade inflation over that period was driven by students choosing  subjects that were more leniently marked, another quarter was driven by improved student quality, but almost half the grade inflation was unexplained and, therefore, likely due to more lenient grading standards over time within the same subject. 
 
The analogy to price inflation is obvious: for the same performance the student receives a higher grade today, compared to what the student would have received a year ago. This is just like consumers paying a higher price for the same product today compared to what they paid a year ago. However, the analogy is not perfect: There is no upper limit on how much nominal prices can increase. Grades, on the other hand, can only increase until everyone gets the best grade. Thus, the term \emph{grade compression} has been suggested as a replacement for the term  \emph{grade inflation} \citep{babcock2010real}. However, the term \emph{grade compression} is silent about the direction of the change in the grading. It could, in principle, be compression in the lower part of the grade distribution. Furthermore, the theoretical implications of grade inflation and grade compression may be very different \citep{zubrickas2015optimal}. Finally, grade compression is, to some extent, studied separately in terms of importance between finer and coarser grading systems. 

Why does grade inflation arise? One source of grade inflation is the  incentives at the individual assessor level to give a generous grade. Individual students are happy to receive good grades because those grades may help them gain admission into a good educational program or career and reflect a positive signal of their human capital to future employers.  The individual teacher and school will also view good grades positively because they can reflect well on the teacher's instruction and may even have a direct financial impact on them.  Moreover, the funding of the school or the salary of the teacher  might be tied to student performance, either directly through accountability systems or indirectly by attracting future students \citep{diamond2016long,nordin2019impact}. In line with these predictions, \citet{wikstrom2005grade} showed that in Sweden, grade inflation is higher in municipalities with more competition between schools.

\citet{chan2007signaling} formalized a signaling theory of grade inflation and showed that schools with many good students have an incentive to apply a more lenient grading policy to help their few mediocre students. However, employers have rational expectations, and if a school has too many weak students,  the signaling value of good grades gets diluted.  Moreover,  \citet{babcock2010real} showed that inflationary grading policies are strategic complements: if one school applies an inflationary policy, other schools will have little difficulty following them, and as a result, grade inflation becomes
contagious. \citet{ehlers2016honest} extended the model of \citet{chan2007signaling} by adding a reputation aspect: after applying a generous grading policy, the next cohort suffers because employers have learned about the lenient grading policy and  will consider this when evaluating future graduates from that school. They showed that this dampens the degree of grade inflation.  \citet{zubrickas2015optimal} formally modeled the grading policy on the student-teacher interaction level. In his model, teachers use grades to incentivize student effort. By applying a more lenient grading policy, the teacher can motivate students farther down the grading distribution at the cost of those at the top of the distribution. In the dynamic version of the model, employers learn about this and realize that that top grades are less informative than they initially thought, which, in turn, gives the teacher an incentive to give the best grade even farther down the ability distribution.  While most theories predict that high ability students tend to bear the individual costs of grade inflation, there are some exceptions. \citet{schwager2012grade} set up a model, according to which  employers use signals other than grades if grade inflation makes the signaling value of grades too weak. As a consequence, they use social class as a signal of ability, which is, of course,  especially costly for high ability workers from disadvantaged backgrounds. The winners are high achieving students from the higher class, because they avoid the competition from high achieving students from the lower class. 

Why is grade inflation an issue of concern? On an individual level the theories on optimal grading policies predict that it harms the high skilled students \citep{babcock2010real,zubrickas2015optimal}. Assessing the individual consequences of grade inflation, \citet{nordin2019impact} studied the impact of institution level grade inflation in Swedish upper secondary schooling. The authors found that students attending institutions with high rates of grade inflation have higher earnings later in life, driven by their post-secondary schooling choices. However, this comes at the cost of high achieving students at institutions that do not apply such high rates of grade inflation. An unusual type of grade inflation happened in France in  1968 as a consequence of a student protest. The protest interrupted the exam period, and as a consequence, passing thresholds were lowered. Students at the margin of passing benefited from this treatment by gaining access to educational programs to which they otherwise would not have access. The students at the margin also received higher wages later in life and their offspring also had better educational outcomes \citep{maurin2008vive}.  While the individual students clearly benefited from more lenient marking in this case, it may very well be that these outcomes would be very different if this policy was applied permanently. For example in this case, the expectation of an exam still gave the students an incentive to learn. If exams were cancelled permanently, this would not be the case. Another potential consequence of grade inflation is that it may affect major program of study and drop-out decisions. \citet{ahn2019equilibrium}  developed a model in which the higher level of observed study effort by women is due to higher perceived benefits of grades. If some programs or institutions have less lenient grading policies, which they document for STEM degrees, women may be discouraged  from entering these fields. \citet{bar2009grade} studied the consequences of the "Cornell experiment" in the mid 1990s when Cornell started to publish median grades for courses. The authors reported that this led to an increased demand for the courses with higher median grades, which led to a higher overall grade point average and grade inflation, in line with the findings of \citet{hernandez2016measuring} discussed previously and also in line with the findings of \citet{sabot1991grade}. The move to the more leniently graded courses was  mainly driven by weaker students. Another cost of grade inflation is that if teachers are very lenient, students may not have a strong incentive to study hard, so they may  reduce their study effort and as a result learn less. Indeed, \citep{babcock2010real} found a negative association between the expected mark (leniency) and study effort.  This correlation is in line with the empirical finding that higher grading standards lead to more learning, especially for high ability students \citep{betts2003impact,figlio2004high}. At the aggregate level, the grade inflation may have important welfare costs in terms of worse matches on the labor because of the noisy signaling value \citep{fredriksson2018mismatch}.  

Are there any benefits to grade inflation? The main cost is the distorted and noisy signal, which may  lead employers to disregard the value of the signal.  \citet{boleslavsky2015grading}  modeled grading standards and school quality in a setting where investment in school quality is endogenous. As a consequence of more lenient grading standards, schools have an incentive to invest more in quality to improve the distinction of their graduates. Although students might face an incentive to shirk under more lenient grading policies, \citet{boleslavsky2015grading} showed that these moral hazard concerns are dominated by the students' anticipation of higher investments by the school.

\subsection{Biases}
At least two sources of bias can be found  in  assessments. First, the design of the  assessment and questions asked may favour some groups of students. Second, the evaluation of the performance in the assessment might be consciously or unconsciously biased towards subgroups of the population. For example, if a girl receives a lower mark for the same performance in an assessment compared to a boy, this would be classified as a bias discriminating against women. The second source of bias especially, including is origin and consequences, has received the attention of economists over the last two decades: the main findings are reviewed next.

Blind assessments, in which  the performance of the student is evaluated without any knowledge about the individual student's characteristics, cannot be biased with respect to the background of the student. However, if the assessment is non-blind, such that the assessor of the performance either observes and possibly knows the student or  can infer the name, gender, or race of the student, the feedback may be influenced by that information. For example, teachers may be more lenient in grading the completed exams of students who typically do well on tests. Alternatively, teachers who give feedback to hundreds of students at the end of the term, who may have difficulty remembering how well individual students did on prior exams, may unconsciously incorporate the group average into the assessment.  It is difficult to remember how well each individual student did, and the teacher might unconsciously include group average in the assessment. For example, if boys historically have done well in mathematics, all boys may receive a slightly higher mark, as a result of stereotyping behavior or statistical discrimination \citep{phelps1972statistical}.

The observation that blind assessments per definition do not include any biases with respect to student background has  been used to empirically test for  evidence of bias in non-blind assessments. The approach was to exploit the fact that students often are exposed to both types of assessments in the same subject and that any systematic differences between these assessments may point toward a gender bias in the spirit of assessments of gender discrimination in orchestra auditions \citep{goldin2000orchestrating}. Alternative explanations exist for the  systematic differences between non-blind and blind assessments, which typically also differ on other dimensions. For example, the blind assessment may be a written and centrally assessed exam, the results of which are then compared to the results of a blind assessment of the classroom performance by the teacher. However, the literature has gone a long way to rule out these alternative explanations for systematic differences between blind and non-blind assessments. Among the first studies to show evidence of systematic differences in assessments was \citet{LAVY20082083}, who compared two scores of high school exams in Israel.  The first score was based on externally and anonymously assessed state exams. The second score was from school internal exams that were assessed by the student's own teacher. Comparing these two scores, \citeauthor{LAVY20082083} found evidence of bias against male students, which increased the gender gap, as girls already outperform boys. Interestingly,  \citeauthor{LAVY20082083} found that this bias varied by teacher characteristics and there is little evidence of stereotyping or statistical discrimination.  

In England, \citet{burgess2013test} showed that  black Caribbean pupils were more likely to receive a poorer assessment by their teacher relative to the blind final test score, than their white peers. In contrast to \citet{LAVY20082083}, these authors found empirical support that indicated  stereotyping was the most likely reason some groups systematically receive more lenient teacher assessments. More specifically, if students belonged to a group that usually did well at a specific school, they were more likely to receive a more optimistic teacher assessment \citep{burgess2013test}.

Similar to the findings from Israel \citep{LAVY20082083}, \citet{falch2013educational} found that Norwegian teachers gave higher marks for girls across all subjects, compared to the blind exam scores, at the end of compulsory schooling. Interestingly, \citeauthor{falch2013educational} studied the hypothesis that the difference was driven by girls doing worse than boys in competitive environments as suggested by laboratory experiments \citep{gneezy2003performance}. The authors exploited that in some regions the results impacted enrollment in subsequent schooling, and in others it did not. However, they concluded that the competitiveness could not explain the gender gap between blind and nonblind grades. Instead, like in the Israeli context, they found suggestive evidence that the gap was  related to teacher characteristics.

Studying the context of the last year of compulsory schooling in Denmark, \citet{rangvid2015systematic} discovered that boys, children whose parents have lower educational background,  and migrants received  lower teacher scores than girls, children of highly educated parents, and natives, with similar blind exam performance. The largest difference was found by parental background, and the smallest difference was by found to be by origin. Studying high school students in Denmark and comparing teacher assessments to blind written assessments, \citet{burgess2022importance} find that compared to girls,  boys were more generously marked by their teacher in mathematics, but less generously marked by their teacher in Danish. 

In the context of 16-year-old Swedish students, Lindahl (2016) investigated whether differences between teacher assessments and national exam results related to the teacher and student being the same gender or sharing a minority status. Contrary to expectations based on evidence from educational literature in general \citep{dee2005teacher}, \citeauthor{lindahl2016teacher} concluded that the differences for girls and non-native students declines when the teacher was female and of non-native origin, respectively. The author interpreted these findings as an indication that teachers were trying to actively compensate for having a  similar background to  students and the potential (unconscious) biases that might involve. 

\citet{terrier2020boys} compared blind and non-blind assessments for middle school children in France and showed that girls received more generous marks from their teacher in mathematics but not in literacy. Finally, \citet{lavy2019persistency} found  large variation in the gender gap between blind and non-blind assessments across subjects for high school students in Greece. 

A particularity of the university system in the United Kingdom is that students apply to programs based on their predicted exam grades, and they receive offers that are either unconditional or conditional on actual exam grades. These predictions are made by their teachers, and any potential biases in predictions may therefore have long-term consequences for these students. Indeed, \citet{murphy2020minority} showed that only 16 percent of the grades were accurately predicted. Most of the teacher predictions were too optimistic, with 75 percent of the predictions being higher than the actual exam result.  A finding of concern reported by the authors was the existence of systematic differences by student background and school type. High achieving low socioeconomic students received a teacher prediction that was  considerably lower than that of their high achieving peers that are not from low socioeconomic households. At the same time, high achieving students in state schools received less optimistic teacher predictions than high achieving students in private schools.  In addition to teacher bias, \citeauthor{murphy2020minority}  listed differences in responses to predictions or a different trend in achievement growth as explanations for these patterns and the showed that these predictions have implications for university application behavior. This study by \citet{murphy2020minority} is slightly different to the earlier cited papers comparing blind and non-blind assessments, as they study differences in predicted results and actual results. One important difference is that these predictions might be intentionally optimistic to encourage student investments and improve student confidence. However, any systematic differences between predicted grades and actual results may come from the same source as in previously cited papers. 

A few studies have applied alternative approaches to identify biases in non-blind assessments. \citet{hanna2012discrimination} conducted an experiment  in India  in which they randomly varied the characteristics of children on their exam sheets before they were marked by local teachers. They found no evidence of discrimination on the basis of age or gender, but discrimination against children from a lower caste was observed. \citet{rangvid2015systematic} exploits a flaw in a grading reform whereby names were replaced by student identifiers to ensure an unbiased assessment. Because the student identifier includes the first four letters of the students' names, this made the gender identifiable for some names but not for others. The study found no clear evidence of significant gender differences between the scripts in which the gender could be identified and the scripts for which  the gender was not identifiable. 

To summarise, evidence has verified systematic gender differences between blind and non-blind assessments favouring girls from Israel, Norway, France, Denmark, Greece, United Kingdom, and Sweden. This difference varied across subjects.  Similarly, evidence has uncovered systematic differences between blind and non-blind assessments that favors  natives and work against students with immigrant status in England and Denmark. Finally,  evidence from Denmark,  India, and the United Kingdom suggests that students from disadvantaged backgrounds are less generously marked by their teachers.  The consequences of these systematic differences between external and internal assessments are discussed later. 

Notably, in most of the studies cited, standardized blind tests were used as the unbiased benchmark. However, statistics showing mean differences in standardized blind tests by gender, socioeconomic background, and origin are often used as a basis to hypothesize that these assessments also suffer from systematic biases. While biases cannot be inferred from such mean differences,  disproving that these assessments are biased is also difficult. One imperfect approach to study this question is to assess the correlation between blind test scores and other (real ) outcomes. If the assessment systematically discriminates against a group, the correlations would under-predict the outcomes for this group. Another strategy is to compare the change in the explanatory power of test scores with and without controlling for some of these group characteristics. Ideally,  the predictive power should be unaffected by adding controls for student background. All these tests are far from perfect, but \citet{sackett2018eight} discussed the various findings and approaches found in research on biases by gender, socioeconomic background, and origin and concluded that the "overwhelming conclusion across decades of research is that tests are not biased against women and racial/ethnic minority group members in their use in predicting subsequent academic performance" and that "there is now extensive evidence countering both the claim that admissions tests measure nothing but SES and the claim that the apparent predictive power of admissions tests is an artifact that disappears when SES is controlled" \citep{sackett2018eight}.

\subsection{Manipulation and cheating}
Assessment outcomes carry important consequences for the students who may not be able to enroll in the desired educational program if their test score does not meet the minimum required for enrollment, for the teacher who may miss out on a bonus or even lose a job, and for the school that might lose funding as a consequence of unsatisfactory student performance or may simply face lower demand from prospective students. These high stakes mean that students, teachers, and the school have strong incentives to  cheat or manipulate test scores. Research has found evidence of such a response,  ranging from "soft" manipulation like targeting the teaching toward what is tested ("teaching to the test") to more direct evidence of cheating and manipulation by actively retrospectively changing the answers on students' completed exams. 

Linking assessments closely to rewards in accountability systems provides strong incentives to improve performance in the subjects included in the accountability system, but this may then come at the cost of students' performance in non-tested subjects or grades \citep{holmstrom1991multitask}. Studying the accountability system in Chicago Public Schools in the mid 1990s, \citet{jacob2005accountability} showed that the system led to improvements in math and reading achievement compared to  earlier trends and compared to  development in other regions. However, non-tested subjects and cohorts saw no improvements, and improvement was mainly driven by test-specific skills and student effort on the test. As noted previously, focusing effort on subjects that are tested and  on specific  test-taking skills is called teaching to the test. Whether strong incentives that lead to teaching  test are bad for learning depends on the costs of monitoring learning and the cost of learning and on the design of the test. If these costs are very high, strong incentives that lead to teaching to the test may still be the most efficient solution \citep{lazear2006speeding}.  Likewise, if the test is well-designed and covers important aspects, focusing on these topics might be worth it. 

Closely related to teaching to the test is the concept of teaching to the rating. Accountability systems may lead to incentives to improve the test scores of certain students. For example, if the accountability is linked to average student performance, the school has an incentive to target the teaching toward students associated with the lowest marginal cost for  improvements to assessment results. In some cases, such as in the No Child Left Behind policy in the United States, the accountability system is linked to several cutoffs in indicators that incentivize targeting resources to students who score right around these cutoffs. \citet{reback} used data from Texas to provide evidence of such behavior. Importantly, they showed that  targeting the resources to students around incentivized cutoffs improves their performance, but that this happens at the cost of students at other parts of the distribution who improve less than they otherwise would. This type of response is closely linked to the concept of "cream-skimming" \citep{heckman2002performance} and multi-tasking where agents focus on the task that is measured at the expense of other tasks that are not measured or counted  \citep{holmstrom1991multitask}. Related to teaching to the rating and teaching to the test, \citet{jacob2005accountability} also finds evidence of the strategic placement of students in terms of placing students in special needs programs to avoid including their performance in the accountability system.

Teachers often face strong incentives to manipulate their students' assessment results, and various approaches to detecting manipulation have uncovered solid evidence that teachers do engage in such practices.  For instance, \citet{jacob2003rotten} studied unusual blocks of correct answers in Chicago Public Schools in the 1990s to exploit that the easiest way for a teacher to change test scores is to adjust blocks of answers rather than randomly adjusting answers for individual students. All students in one classroom having identical, correct answers to a set of questions, such as 8 through 15, but different answers to all other questions would be an indication of such a manipulation, especially if the blocks of correct answers vary from classroom to classroom. Also, evidence of students in a classroom experiencing unusual test score gains, with their scores suddenly improving far beyond what their regular trajectory would predict and then returning to their old trajectory, was also considered an indication of manipulation among the school children in Chicago \citep{jacob2003rotten}. Another sign of manipulation is unusual test score distributions that is aligned with teacher or school incentives. For example a test score distribution that shows unusual low mass below incentive cutoffs and unusual high mass just above incentive cutoffs,  suggest that their results were pushed above the thresholds. This approach was used to detect teacher manipulation in \citet{diamond2016long} and \citet{dee2019causes}.

Using the first two strategies described, \citet{jacob2003rotten} uncovered evidence of test score manipulation in four to five percent of all classrooms in Chicago Public schools. As these strategies rely on unsophisticated manipulation, they are easy to detect using statistical rules. In the setting of the New York State Regents examinations, teachers locally mark their own students' exams. \citet{dee2019causes} found that the test score distributions in the Regents exams exhibited discontinuities in line with students being pushed above important cutoffs in the marking. Such sharp discontinuities are unlikely to happen due to teaching to the rating. The authors further revealed that more than 40 percent of scores below important cutoffs were manipulated and moved above the cutoffs, accounting for six percent of all tests. \citeauthor{dee2019causes} also showed that this type of manipulation disappeared once the marking became part  a centralised system. Manipulation by  teachers has important consequences for students who have a greater chance chances of graduating from high school. However, the manipulation also has harmful effects  as some students do not complete advanced courses as a consequence of being manipulated above a cutoff. This points to the importance of grades for updating beliefs about investments and skills. Students who are pushed above a threshold might be misled to believe that their understanding of the material is better than it actually is, and they might therefore not invest sufficiently in terms of subsequent study effort. 

Research shows that teacher manipulation also occurs in settings with softer incentives than in the New York setting studied by \citet{dee2019causes}. For example, in Sweden, where no direct incentives to manipulate test scores exist, \citet{diamond2016long} show evidence of test score manipulation in high schools. The incentive for the schools and teachers in this setting was mainly to attract future students by documenting high passing rates and high grade point averages. While  test score manipulation for financial reasons may go against the teacher's professional norms, the authors demonstrated that the test score manipulation mainly occur for students who do worse than predicted given prior performance. They did not find any evidence that teachers were more likely to manipulate the performance of students with certain background characteristics.  In other words,  teachers appear to be  trying to correct for  a bad test day and for potentially being exposed to the negative exogenous factors I discuss below. As in the case of the United States, test score manipulation has long-term consequences for the students in Sweden. The students who had their grades manipulated upward were more likely to enroll in a university degree program, complete more years of education, were less likely to have a child as a teenager, and had higher earnings at age 23. 

\citet{lin2020catching} investigated student  cheating on exams by comparing the number of identical answers for neighboring students to the number of identical answers for students sitting farther away. They found that when students sat next to each other,  they had more shared
incorrect answers and  more shared correct answers, compared to if two students were  sitting more than  than two seats apart. The increase was only found among students sitting directly next to each other. There was no significant increase in the number of shared answers for students sitting behind each other or farther than two seats apart.

\subsection{External factors}
Most assessment situations involve a number of factors that affect student performance that are outside the control of the student and often also outside the control of the  teacher or school \citep{kane2002promise}.  Such external factors, like  temperature and air quality, tiredness, and well-being at the time of the assessment may affect students' condition and ability to perform well.  Evidence of the impact of some of these factors is provided in this section.

\subsubsection{Air quality and pollution}
Evidence from various settings has shown that individuals' performance is affected by air quality and pollution, for example, impacting productivity on the labor market \citep{graff2012impact} and  the performance of chess players \citep{kunn2019indoor}. Importantly, evidence has also shown that pollution affects students' performance in important assessments. \citet{ebenstein2016long} studied the performance of students completing high-stakes exit exams in Israeli secondary schools, known as the Bagrut. Students' performance on these exams has implications for their chances to enroll in university programs. The authors assessed how air pollution affects the students' performance in these tests by exploiting day-to-day variation in PM$_{2.5}$ levels due to forest fires and sandstorms. Because the students sit for the Bagrut across several days and  sit for several tests, the authors were able to compare performance across different levels of pollution exposure for the same students. Importantly, the students could self-select into an exam time and, instead follow a pre-determined schedule  unrelated to individual and school characteristics.  The main finding was that a one standard deviation increase in air pollution caused a four percent of a standard deviation drop in performance, with larger effects for boys, low-performing students and students from disadvantaged backgrounds. As these exams are important for subsequent university enrollment, the pollution also impacted later life outcomes. Indeed, \citet{ebenstein2016long} found that higher pollution levels at the time of the test causes lower levels of completed education and lower earnings. 

\subsubsection{Tiredness}
Individual health and tiredness at the time of the assessment is likely to also affect  performance. However, obtaining accurate measures of health and tiredness at the time of assessment to enable these factors to be studied is very difficult, as is finding variations in these factors that are not related to other student characteristics. For example, students from disadvantaged backgrounds may, on average, be  more tired at the time of the assessment because of lower quality housing than their peers from other households, and this variation in tiredness will then be  confounded by other factors that vary across these students, such as general educational support and nutrition. To circumvent these issues and still shed light on the importance of tiredness, \citet{sievertsen2016cognitive} exploited variation in  time of the day the individual was assessed. The authors relied on the fact that the timing of the test depends on a combination of the regular schedule and the school level computer availability and  is, thus, unrelated to individual characteristics. Similar to \citet{ebenstein2016long}, \citet{sievertsen2016cognitive} compared the test performance of the same individual sitting for several tests and found that for every hour later in the day, that the test was taken the test performance decreases by 0.9 percent of a standard deviation. Interestingly, the authors found that breaks can mitigate these effects.  

\subsubsection{Distractions}
A dog barking outside the testing room or neighbouring students making distracting noises will likely also affect student performance on tests; however, there is limited evidence on  causal effects of such distractions. One study has, however, studied the effect of wider distractions in terms of the value of leisure around the test-taking and test preparation time. \citet{metcalfe2019students} investigated the effect of having assessments in times of large football tournaments, such as the European Championships and the World Cup, for  students in England sitting the General Certificate in Secondary Education or GCSE achievement test. These tournaments are very popular and involve broadcasting of matches, parties, and other distractions that are likely to raise the value of leisure and increase the opportunity costs of studying for exams. The authors uncovered that exposure to this distraction mainly is driven by year of birth, as these tournaments only take place every other year, and found that students exposed to this distraction do worse in the assessments. The effects are largest for boys from disadvantaged backgrounds. While the study by \citet{metcalfe2019students} provides insights into how distractions affect learning in general, given that the distraction is centered around assessment, it also points to the importance of the environment around the assessments. 

\subsubsection{Stress and nutrition}
While general stress levels and nutrition quality rarely are considered as exogenous in the sense that they are correlated with parental background, stress levels and nutrition quality are often outside the control of the individual child. \citet{heissel2021testing} show that children in poorer neighbourhoods have higher stress levels and that higher levels of stress are associated with lower test scores.  Studying the impact of nutrition and food, \citet{bond2021hungry} investigated the as good as random timing of the assignment of the Supplemental Nutrition Assistance Program (SNAP) benefits in the United States. The authors exploited that in some states the monthly timing of the benefits depends on the first letter of the surname, combined with information on names with the timing of SAT college admission exam. The results show that students who attend this high-stakes assessment just before they would receive new benefits (i.e., at the time of highest food insecurity) score six percent of a standard deviation lower on the exam, compared to those sitting for the exam just after receiving the benefits. Because the SAT results are important for university acceptance, these results carry over into a lower probability of attending a four-year university program for the low-income students. 

There is also evidence of systematic differences in how students respond to the stakes of assessments. \citet{azmat2016gender} showed that girls outperformed boys in tests, but the gap declined as the stakes of the test increased. One possible explanation for this pattern is that boys exhibit low test effort when stakes are low. An alternative explanation is that girls perform worse under pressure than boys. One piece of evidence going against the first explanation is that girls actually perform worse on low stakes tests. The authors hypothesized, therefore, that differences were due to gender differences in response to pressure \citep{azmat2016gender}.

\section{Assessments and individual behavior and outcomes}
How do assessments affect the individual? Assessments provide information  about the learning progress. The individual might use this information to update beliefs about the returns on study effort and beliefs about their own abilities. Assessments also provide incentives. A good result in an exam may provide easier access to the desired educational program, a scholarship, or a dream job. Assessments, therefore, often also provide an incentive for the student to invest in study effort and maybe select courses in which they know they will get a better mark, which affects their learning and human capital development directly.  In this section,   research  on how assessments affect individual behavior and outcomes is discussed. 

\subsection{Learning }
A return to the initial discussion on assessment types and objectives in this chapter will be useful to the discussion on how assessments affect learning. Standardized testing is often introduced in connection to accountability systems. The contrasts to these schemes are internal tests that are designed and conducted by the teacher. While the former has the advantage of producing results that are comparable across classrooms, schools, and maybe countries, the latter has the advantage of being more targeted toward the specific classroom teaching.  While  pedagogical research points to standardized tests being ineffective for (or even detrimental to) learning, because these assessments provide no detailed feedback that the teacher and student can use to improve  learning and because the assessment is not tailored to what is taught in that classroom \citep{guskey2007using}, still  such assessments might also affect learning by acting as  incentives for students, teachers, and schools to invest effort. The incentive mechanism is closely linked to the information mechanism. Assessments provide information to teachers and students about their progress. The students (or the parents) can  learn about the returns on study effort and adjust effort levels accordingly or change study strategies. Likewise, teachers might adjust teaching methods or target specific areas of the curriculum following an assessment. 

Looking at the big picture,  \citet{bergbauer2021testing} compared  PISA attainment data for two million students in 59 countries over the period 2000 to 2015 together with information about the use of assessment systems in the schools. Based on a cross-country panel-regression exploiting changes across time within countries, but controlling for aggregate time trends and country level differences, they found that standardized testing raised attainment in PISA test scores. Importantly, they found that the benefit of testing is seen in low-performing countries, whereas they found no evidence of a significant association between standardized testing and PISA results  in countries that already do well in the PISA assessment. Another important finding of \citet{bergbauer2021testing} is that they only found a significant positive effect from standardized testing, but no evidence  of any impact of internal  testing and reporting. One threat to these conclusions is that national standardized testing might be introduced simultaneously with other policies that can explain part of the effect. \citet{bergbauer2021testing}  ruled out and control for the impact of number of  alternative policies and explanations; their findings were also supported by a micro-level study focusing on an exogenous shock to testing in Denmark \citep{andersen2020learning}. In this study, the authors investigated the subsequent performance of students who had missed sitting for a national standardized assessment in Danish schools due to an IT crash. They discovered that these students performed worse on tests in subsequent years, an effect that was stronger for students for disadvantaged backgrounds. One downside of these approaches is that the outcome is closely related to the treatment. Students may do worse in subsequent assessments because they had less practice in taking tests. Future research can link the absence of testing to other outcomes as well.  

It is not entirely clear how the assessments affected learning in the case of \citet{bergbauer2021testing} and \citet{andersen2020learning}. One potential mechanism is that not completing a test means that students miss out on a chance to get (unbiased) feedback on their learning. In the context of \citet{andersen2020learning}, the next assessment was three years after the IT-crash, illustrating that a missed opportunity for feedback and reacting to this feedback might mean three years of not getting the extra support needed or not making required changes to study strategies. To shed  light on how test feedback promotes learning  \citet{beuchert2020impact} identified the effect of scoring just below a cutoff (e.g., scoring  "Below average" instead of "Average"). The authors found that children receiving negative feedback in mathematics do better in subsequent maths assessment compared to students who only did marginally better at the initial assessment. They also found evidence of a positive spillover to other subjects. These findings suggest that  test feedback is used actively and affects future learning. 

As previously discussed,  evidence from several countries has shown that non-blind teacher assessments differ systematically from blind assessments. A number of the aforementioned studies documenting these biases also assess the consequences for students. \citet{terrier2020boys} calculated  individual teacher favoritism for girls and exploited variation in this measure, and under the assumption of the random assignment of teachers to students in sixth grade, she demonstrated that this causes a widening gender learning gap in mathematics and French. In mathematics this widening was driven by girls performing better (and boys not being affected) and in French the gap was driven by boys doing worse (and girls not being affected). Furthermore, for girls, being assigned to a teacher who favours girls  increases the likelihood of choosing a science track in high school four years after being exposed to that teacher showing favoritism towards girls. Similar conclusions have been reached in  other settings; for example, \citet{lavy2019persistency} provided evidence that assessments matter for subsequent learning, potentially through affecting students' expectations and subsequent investments in study effort.  These findings are in line with the findings of \citet{burgess2022importance}  that indicated that a semi-blind assessment in mathematics at the end of high school reduces the gender gap in enrolling and graduating in subsequent university STEM degrees. 

 A growing area of the literature addresses ways that assessments motivate student learning, especially if the result of the assessment has  important consequences for the individual.    \citet{hvidman2021high} examined a grading reform that required high-stakes grades to recoded. As  students knew that the change in their grades was due to these recodings and not due to their individual characteristics or performance, the only reason for them to react to this reform was that the grade point average was important for them. \citet{hvidman2021high} showed that students who were negatively affected did better in subsequent assessments to compensate for the negative recoding. This reaction had long-term effects on their enrollment in subsequent educational programs. Interestingly, the response was driven by girls. There was no response in the short run or in the long run by boys. Another setting that provided evidence of grades acting as incentives to increase learning is Greece, where the chance to retake an important high-stakes assessment was shown to improve learning considerably \citep{bizopoulou2022second}. Somewhat contradicting these strong incentive responses, a study on university students in Texas found no evidence that students who were close to receiving a better letter grade (A-F) did better than students who were farther away, despite their stronger incentive \citep{grant2013grades}, which might be explained by differences in the importance of the assessment.

\subsection{Belief updating, course choice, and drop-out decisions}
\subsubsection{Belief updating and choices}
Beyond using assessments to adjust study effort, students can use them to update their beliefs about their skills. 
\citet{stinebrickner2012learning} studied how students in a Canadian university  used feedback from assessments in their decisions about study effort and drop-out decisions.  They found that at university entry, students were too optimistic about how they would do, which was mostly explained by being too optimistic about their own abilities (in contrast to being too optimistic  about their study effort). However, students then updated their beliefs actively based on assessment results, with some heterogeneity depending on views of explanations for assessment results. The take-away message is that 40 percent of all university drop-outs result from students' learning about their abilities from assessments.  Studying the same setting on university students in Canada, but focusing on major choice, \citet{stinebrickner2011math} found that grades play an important role in explaining why fewer students major in STEM subjects than beliefs at university entry suggest, in line with the previously  discussed findings by  \citep{ahn2019equilibrium} related to how gender differences in perceived benefits of grades possibly  explaining gender differences in pursuing and completing STEM majors.  On the other hand, \citet{andrew2011adoption} documented limited evidence of updating expectations and beliefs in response to changes in grade point averages. Responses were modest and only in reaction to large changes. The topics of belief formation and expectations in  education, including the role of feedback from assessments, has been studied by sociologists for decades.

Grades  affect individuals' choices through information about the individuals' own abilities but also by giving them incentives to attend courses that are more leniently assessed if they care about their overall grade point average, as discussed in the section on grade inflation \citep{bar2009grade,ahn2019equilibrium}.

\subsubsection{Relative feedback}
How students use  assessments as information to update their beliefs depends  on both their prior beliefs and how the feedback is provided. One key aspect of the latter is whether the assessment feedback provides information on the relative performance. To study the importance of relative feedback, \citet{azmat2019you} carried out an experiment in a Spanish classroom where students in the control group only received information about their own performance, while students in the treatment group also received information about their relative performance. In subsequent assessments, students in the treatment group were less likely to pass, and they received lower grades. Using survey data the authors also show that the treated students were more satisfied. The response to relative performance information is agrees with a theory that students, on average, underestimate their relative rank. This suggests that students care not only about their absolute scores, but also about their relative rankings. 

\subsubsection{Teacher bias and student outcomes}
As discussed previously, \citet{terrier2020boys} showed that teacher favoritism in non-blind assessments affect high school track choice. This was also confirmed by \citet{burgess2022importance}, who showed that semi-blind assessments in mathematics at the end of high school reduced the gender gap in enrolling and graduating from university STEM degree programs. The counter-factual treatment would be a semi-blind assessment in another subject (like Danish) and mainly a non-blind teacher assessment in mathematics. They also all sit for a fully blind mathematics exam. As they show that teachers are more lenient to boys in the their assessments, compared to the exams, this points to the importance of assessments for track choice. 

Students who were under-predicted in the United Kingdom \citep{murphy2020minority} were also less likely to apply to a high-tier university and more likely to be overqualified for the programs to which they apply. Taken together, the systematic differences in prediction error by student background and the implications of the prediction error for university application behavior suggests that the system of using predicted grades for university applications may have implications for getting high achieving students from disadvantaged backgrounds matched with appropriate university programs.

\subsubsection{Heterogeneous belief updating}
Some evidence of gender differences in response to grades and grading policies has been reported, which explains part of the gender difference in enrolling and majoring in STEM degree programs \citep{ahn2019equilibrium}.  At the aggregate level, \citet{bergbauer2021testing} concluded that increased standardized testing is especially useful in low-performing countries. Likewise, at the micro level, \citet{andersen2020learning}  uncovered evidence that children from disadvantaged backgrounds suffered the most from not having access to standardized testing. On the other hand, \citet{beuchert2020impact}  found no difference in the response to test feedback by parental background. While there is evidence that students from disadvantaged backgrounds have lower expectations about their future educational careers than students from more affluent backgrounds who do equally well in assessments, \citet{karlson2019expectation} found no evidence that students from disadvantaged backgrounds were less responsive in their updating of expectations. 

\subsubsection{Rank effects}
A rapidly growing literature has examined the role of ordinal rank in the classroom setting. These ranks are often determined based on assessments. There is solid evidence that students' relative rank in the classroom affects their later outcomes.  Exploiting the university setting of random assignment to peer groups, \citet{elsner2021achievement} showed that a higher rank increased performance. In other words, a student with a given ability will do better if the student is randomly assigned to a peer group with students ranked lower than the student  than with students who rank higher than that student. This rank  effect is also found to impact risky behavior (unprotected sex and engaging in physical fights) \citep{elsner2018rank}, subsequent educational choices and investments \citep{elsner2017big,murphy2020top}, and earnings \citep{denning2018class}.

\subsection{Mental health and well-being}
As discussed, the importance of an assessment for future outcomes can motivate a student to study harder and learn more; it may also have negative consequences on the  well-being of the individual. Using data on children aged eight to 15 in New Orleans, \citet{heissel2021testing} studied the link between high-stakes testing and stress levels. The authors documented 18 percent higher levels of saliva-based measures of cortisol, a measure of the stress hormone capturing how the body's stress system is functioning, in weeks with high-stakes testing (before taking the test) compared to the same time in non-testing weeks. The effect was mainly driven by boys, and while the   sample was relatively disadvantaged, they also found suggestive evidence that the response was largest for children living in poorer neighbourhoods. Importantly, \citeauthor{heissel2021testing} also showed that large responses in stress levels were associated with worse performance. 

In Germany, children are assigned to an academic or non-academic track around the age of ten to 11. \citet{bach2020understanding} studied how well children do in schools where the track recommendation is linked to the performance on assessments compared to the children in  the schools where the track recommendation were more vague. In line with evidence that the stakes of the assessment incentivize student effort \citep{sievertsen2016cognitive}, \citeauthor{bach2020understanding} discovered that students in the schools where track choice is tightly linked to assessments show stronger growth in assessments. However, they also found that these children had worse mental health and were more likely to be stressed. These results point to an important downside of high-stakes testing.   In the previously mentioned article on test feedback and subsequent performance exploiting discontinuities in feedback, the authors also studied the effect on well-being. While the data was imperfect for this test, the authors found no evidence that scoring just below a cutoff  affects mental health negatively \citep{beuchert2020impact}.

It is worth highlighting that it is more difficult to quantify the impact on mental health than on learning, because the latter, almost by definition, is quantified and measured automatically, and the latter is rarely captured and depends on the availability of survey data. An important area for future research would be the persistence of effects of testing on stress and mental health. 

\subsection{Long-term  outcomes}
\subsubsection{Access to degrees}
Assessments in school and education may affect long-term outcomes of the students through several channels. Firstly, assessment results are used to sort students into subsequent educational programs, and exogenous factors and the design of these assessments will, therefore, affect the educational programs in which the students can enroll. Secondly, assessment results might be used by employers to screen applicants and, as such, will affect the match on the labor market, their earnings, and their employment chances. This section presents the evidence on these two factors. Thirdly, assessments provide students with information about the returns on studying and about their skills. Students may respond to this information by selecting  specific tracks and increase or decrease their study effort. 

As noted previously, evidence on teacher manipulation shows that students who benefited from the manipulation in the sense that their scores were improved because it also (in most cases) had better later life educational outcomes \citep{diamond2016long, dee2019causes}. Likewise, as discussed in the case of more lenient grading standards, the higher grade given was shown to benefit students at the margin of passing who obtained access to degrees they otherwise would not have had access to \citep{maurin2008vive}.  One direct explanation for this is that being pushed above a threshold gives access to more educational programs. This mechanism was confirmed by a study on national exams for all students at the end of compulsory schooling in England \citep{machin2020entry}. Comparing the subsequent educational trajectories,  students in England who barely failed to receive a "good pass" in English in these high-stakes exams were less likely to enroll in upper-secondary academic and vocational tracks and start tertiary education, three years after the exams. 

Prior grades and admission exams are often used to sort students into subsequent tracks and educational programs. For example, in the United States, the ACT or SAT admission exams are required for admission to most four-year colleges. These admission tests were historically taken outside  school testing by students interested in attending college. This gives rise to several sources of inequality. Firstly, students and their parents need to know about this system and should also know how these tests work in order to do well. Secondly, direct costs are  involved in terms of fees paid to the test centers. Thirdly, indirect costs are also involved in terms of travel to the test center outside school days. \citet{hyman2017act} studied the consequences of these admission tests for access to college by exploiting policy variation through the introduction of mandatory ACT testing in Michigan in 2007. Following this policy, the testing was moved  the school day at no costs to the students or the schools. This policy should, thus, alleviate all three concerns about inequality: information, direct and indirect costs. The overall response to the policy was a two percent increase in college enrollment, with substantially larger effects in schools with high poverty shares. Importantly, \citeauthor{hyman2017act} found no evidence that the marginal student pushed into college through this policy is likely to graduate from college. A take-away figure from the study is that for every ten poor students who get at least a college ready score on the ACT prior to the policy, another five would have scored above that threshold but did not attend the test before the reform.  Studying the introduction of mandatory SAT testing in Maine, \citet{hurwitz2015maine} revealed similar overall results to \citet{hyman2017act} in a setting where the tests took place outside the school schedule but at no cost and with transport support.

Further evidence that financial costs constitute a barrier for low-income students is provided by \citet{goodman2020take}, who found that low-income students were less likely to retake the SAT university admissions test, but that low-income students who have waivers for test fees have higher retake test rates than their low-income peers without waivers. In this study, the authors observed that students were more likely to retake the SAT test if they were close to a score that was a multiple of 100. In other words, students were significantly more likely to retake if they scored 2099 than if scored 2100. Importantly, retaking the test caused higher scores, and lower retake rates among low-income students may, therefore, be another source of inequality, especially because \citet{goodman2020take} showed that retaking was especially beneficial for disadvantaged students.

\subsubsection{Job-market signaling}
Under the assumption that individuals with higher labor market productivity also have lower costs from completing educational degrees, the job-market signaling model suggests that employers use years of completed education to screen workers \citep{arrow1973higher,spence1978job}. The key insight is that workers who have completed more education will receive a higher wage, not necessarily because the education made them more productive but simply because the education allows employers to extract information about the individuals' innate productivity.  It is straightforward to extend the job-market signaling model from level of education to educational credits in terms of how well the students did in school. If students with higher labor market productivity are assumed to also have lower costs from doing well on assessments, the employers can use assessment results to screen workers. 

Grades may not only affect individuals' choices through information about their own abilities but also by giving them incentives to attend courses that are more leniently assessed if they care about their overall grade point average, as discussed in the section on grade inflation \citep{bar2009grade,ahn2019equilibrium}.
To test whether employers use assessment to sort workers, \citet{piopiunik2020skills} conducted an experiment where they sent fictitious job applications to German employers, with randomly varied elements on the applicants' curriculum vitae, including their grade point average. The researchers reported that a higher grade point average increased the likelihood of a job interview, and that the effect was stronger for college graduates than for high school graduates. They uncovered no differences by applicant gender, but they found that larger firms cared more about the grades than smaller firms and that older human resource managers cared less about grades. The findings by \citet{piopiunik2020skills} aligns with earlier evidence from \citet{koedel2012math} who conducted a similar experiment in the United States, that showed that a signal of high mathematics skills on the curriculum vitae increased the likelihood of a positive response by the employer. In a similar study conducted in Germany, \citet{protsch2015employers} documented evidence that employers use certain grade point average thresholds in their recruitment decisions. 

Looking beyond the experimental variation in fictitious applicants and studying actual job market outcomes, \citet{clark2014signaling} used data on high school students in Texas and compared labor market outcomes of students who barely passed the high school exit exam to those who barely failed the exam. Both groups of students received a certificate, the former a certificate for obtaining the high school degree, and the latter a certificate for completing high school. As students very close to the cutoff should be essentially identical in their skills, any difference in labor market outcomes should reflect the signaling value of actually passing the degree. However, the authors found no evidence of this signal on subsequent earnings. While this finding goes against the findings of the fictitious application studies, it is worth highlighting that \citet{piopiunik2020skills} found that the effects were considerably smaller among high school applicants than college applicants and that \citet{protsch2015employers} found suggestive evidence that employers use specific cutoffs in the grade point averages to screen potential workers.  Using a  design similar to that of \citet{clark2014signaling}, \citet{jepsen2016labor} compared labor market outcomes for individuals barely passing the General Educational Development (GED) certification to the outcomes of individuals who barely failed it. They found no effects on subsequent employment for women but larger earnings in year two after earning the certificate. 

Another possible reason for why the two studies produced weak evidence of the signaling value of grades and passing exams, while the survey experiments suggested large labor market effects, is that these two studies focused on local threshold effects for a rather particular degree type. \citet{hansen2021grades} used a  grading reform to study the effect of as-good-as random variation in grade point averages for graduates from the two largest universities in Denmark. The research exploited the same recoding as in \citet{hvidman2021high} which has the advantage that it affects the entire grade distribution. They found that an exogenous increase in grade point average caused significantly higher earnings in the first years after graduation, but that the effect disappeared within two to four years. This suggests that that employers initially use the grades to screen potential workers, but that they quickly learn about the workers' skills.

\section{Taking stock}
This section summarizes the evidence covered in this paper related to the impact assessments have on efficiency, equity, and well-being but first presents a discussion on the impact of assessments on schools and beyond.
\subsection{Effects on schools}
 Assessments may have a direct impact on the school's finances and on individual teachers' pay and employment  when directly linked to accountability incentives like in case of the NCLB policy in the United States \citep{jacob2005accountability}. Schools might also be indirectly affected by assessments, if assessment results are publicly available and parents and students use this information to decide what school to attend \citep{diamond2016long, nordin2019impact}. As discussed, such incentives might affect schools' and teachers' behavior such as teaching to the test placing students on special needs programs \citep{jacob2005accountability}, teaching to the rating \citep{reback}, teacher test score manipulation \citep{jacob2003rotten,diamond2016long,dee2019causes}, and lenient grading policies \citep{nordin2019impact}. 

There is evidence that school's prioritization of grades based on the expectation that they matter for parents is justified.  In the school choice literature, the evidence that parents value school quality is solid \citep{black1999better, imberman2016does,burgess2015parents, borghans2015parental, harjunen2018best}. School average test scores  are a product of both the quality of schools and the student composition, whereas value added is meant to capture only the school quality. A school that prioritizes doing well on assessment either through improved teaching or strategic responses (e.g., teaching to the test or rating) will improve both raw test scores and value added. To mention a few examples, one strategy to assess the degree to which parents care about test scores, is to study the revealed preferences in terms of their willingness to pay. Looking at house prices in Massachusetts around school district boundaries, \citet{black1999better} concluded that  a five percent increase in school test scores caused an increase in the willingness to pay for housing by 2.1 percent. The boundary idea is based on the premise that houses close to the school district boundary should be essentially identical in terms of all other amenities except for the school, which is allocated based on the district. However, similar effects have also been found in  Finnish settings where the assessment results were not publicly available \citep{harjunen2018best}.  \citet{imberman2016does} used data from Los Angeles and confirmed the finding that assessment results affects house prices. Value added measures, on the other hand, do not, which suggest that parents care about test scores, but maybe less about the origin of test score improvements. There is some evidence for heterogeneity in how parents value school quality. Studying school choice in England, \citet{burgess2015parents} showed that while most parents had preferences for academic performance of the schools, some socioeconomic gradient existed in these preferences. For example, the preference for academic quality was stronger for high socioeconomic parents. However, this may also be due to constraints, either actual access constraints or information constraints. In sum, there is clear evidence that a school's assessment result affect parents' demand and that they are willing to pay for a school that receives better results. Less clear is if  the parents care about whether the good assessment comes directly from school quality, or through better students and, thus, better peers. 

\subsection{Assessments and inequality}
The previous sections clearly showed that assessments may affect inequality through several channels. This paragraph briefly summarises some of the main channels: Firstly, there is evidence that some groups suffer from biases in assessments \citep{LAVY20082083,burgess2013test,falch2013educational,rangvid2015systematic,terrier2020boys}. Such biases may have longer-run effects on students educational outcomes, because students use assessments to update their beliefs about returns to studying and their abilities \citep{stinebrickner2012learning}, which is also directly documented in various settings \citep{terrier2020boys,murphy2020minority,burgess2022importance}. Thirdly, heterogeneity in access to and response to assessments affect test taking behavior and subsequent access to education \citet{hyman2017act,andersen2020learning,goodman2020take}. Fourthly, there is evidence of socioeconomic differences in how students are affected by and exposed to external shocks in test taking \citep{ebenstein2016long}

\subsection{Assessment and efficiency}
The research evidence  also clearly shows that assessments affect efficiency in terms of human capital decisions. The decision of the optimal level (and later type of) schooling has been considered an investment at  least since some of the early models on optimal  human capital decisions \citep{schultz1961investment,becker1964human}. In these models, schooling raises productivity and we invest in it until marginal returns equal the marginal costs. However, as the evidence shows, individuals' educational decisions are affected by the leniency in the assessments \citep{bar2009grade}, likely because employers use the grades as a screening device \citep{piopiunik2020skills,clark2014signaling,protsch2015employers,jepsen2016labor,hansen2021grades}. Moreover, students react to the high-stakes incentives in assessments by increasing their effort \citep{sievertsen2016cognitive}, and such strategic responses are actually also observed by schools \citep{reback} that focus on the students who count more in the accountability system. In sum, because assessments are used as a signaling device  in accountability systems, students' and schools' investment in education might differ from an optimal decision in terms of maximizing learning and, thus, human capital accumulation. 

\subsection{Mental health and well-being}
It is clear from the  evidence discussed, that our knowledge about how testing affects mental health and well-being is relatively limited. One explanation for this is that the outcome, mental health and well-being, is both harder to measure and less frequently measured automatically, compared to, for example, learning and later life outcomes. However, the existing evidence points to effects on stress \citep{heissel2021testing}, and also on mental health in very young ages \citep{bach2020understanding}. Given the strong evidence on biases on how assessments affect incentives to study hard, drop-out decisions, course choices, access to subsequent educational programs, and labor market outcomes, that there may effects on mental health and well-being is not surprising. However, much more research is needed to answer questions on how to balance the benefits of standardized testing for equity \citep{hyman2017act} and  potential detrimental effects on well-being.

\bibliography{library}

@article{bergbauer2021testing,
  title={Testing},
  author={Bergbauer, Annika B and Hanushek, Eric A and Woessmann, Ludger},
  journal={Journal of Human Resources},
  pages={0520--10886R1},
  year={2021},
  publisher={University of Wisconsin Press}
}

@article{ebenstein2016long,
  title={The long-run economic consequences of high-stakes examinations: Evidence from transitory variation in pollution},
  author={Ebenstein, Avraham and Lavy, Victor and Roth, Sefi},
  journal={American Economic Journal: Applied Economics},
  volume={8},
  number={4},
  pages={36--65},
  year={2016}
}

@techreport{hansen2021grades,
  title={Grades and employer learning},
  author={Hansen, Anne Toft and Hvidman, Ulrik and Sievertsen, Hans Henrik},
  year={2021},
  institution={IZA Discussion Papers}
}

@article{jepsen2016labor,
  title={Labor market returns to the GED using regression discontinuity analysis},
  author={Jepsen, Christopher and Mueser, Peter and Troske, Kenneth},
  journal={Journal of Political Economy},
  volume={124},
  number={3},
  pages={621--649},
  year={2016},
  publisher={University of Chicago Press Chicago, IL}
}

@article{protsch2015employers,
  title={How employers use signals of cognitive and noncognitive skills at labour market entry: insights from field experiments},
  author={Protsch, Paula and Solga, Heike},
  journal={European Sociological Review},
  volume={31},
  number={5},
  pages={521--532},
  year={2015},
  publisher={Oxford University Press}
}

@incollection{spence1978job,
  title={Job market signaling},
  author={Spence, Michael},
  booktitle={Uncertainty in economics},
  pages={281--306},
  year={1978},
  publisher={Elsevier}
}

@article{koedel2012math,
  title={Math skills and labor-market outcomes: Evidence from a resume-based field experiment},
  author={Koedel, Cory and Tyhurst, Eric},
  journal={Economics of Education Review},
  volume={31},
  number={1},
  pages={131--140},
  year={2012},
  publisher={Elsevier}
}

@article{piopiunik2020skills,
  title={Skills, signals, and employability: An experimental investigation},
  author={Piopiunik, Marc and Schwerdt, Guido and Simon, Lisa and Woessmann, Ludger},
  journal={European Economic Review},
  volume={123},
  pages={103374},
  year={2020},
  publisher={Elsevier}
}

@article{black1999better,
  title={Do better schools matter? Parental valuation of elementary education},
  author={Black, Sandra E},
  journal={The quarterly journal of economics},
  volume={114},
  number={2},
  pages={577--599},
  year={1999},
  publisher={MIT Press}
}

@article{clark2014signaling,
  title={The signaling value of a high school diploma},
  author={Clark, Damon and Martorell, Paco},
  journal={Journal of Political Economy},
  volume={122},
  number={2},
  pages={282--318},
  year={2014},
  publisher={University of Chicago Press Chicago, IL}
}

@article{gneezy2019measuring,
  title={Measuring success in education: the role of effort on the test itself},
  author={Gneezy, Uri and List, John A and Livingston, Jeffrey A and Qin, Xiangdong and Sadoff, Sally and Xu, Yang},
  journal={American Economic Review: Insights},
  volume={1},
  number={3},
  pages={291--308},
  year={2019}
}

@article{lin2020catching,
  title={Catching cheating students},
  author={Lin, Ming-Jen and Levitt, Steven D},
  journal={Economica},
  volume={87},
  number={348},
  pages={885--900},
  year={2020},
  publisher={Wiley Online Library}
}

@article{jacob2005accountability,
  title={Accountability, incentives and behavior: The impact of high-stakes testing in the Chicago Public Schools},
  author={Jacob, Brian A},
  journal={Journal of Public Economics},
  volume={89},
  number={5-6},
  pages={761--796},
  year={2005},
  publisher={Elsevier}
}

@article{jacob2003rotten,
  title={Rotten apples: An investigation of the prevalence and predictors of teacher cheating},
  author={Jacob, Brian A and Levitt, Steven D},
  journal={The Quarterly Journal of Economics},
  volume={118},
  number={3},
  pages={843--877},
  year={2003},
  publisher={MIT Press}
}

@article{hernandez2016measuring,
  title={Measuring inflation in grades: An application of price indexing to undergraduate grades},
  author={Hernandez-Julian, Rey and Looney, Adam},
  journal={Economics of Education Review},
  volume={55},
  pages={220--232},
  year={2016},
  publisher={Elsevier}
}

@article{wikstrom2005grade,
  title={Grade inflation and school competition: an empirical analysis based on the Swedish upper secondary schools},
  author={Wikstr{\"o}m, Christina and Wikstr{\"o}m, Magnus},
  journal={Economics of education Review},
  volume={24},
  number={3},
  pages={309--322},
  year={2005},
  publisher={Elsevier}
}

@article{bach2020understanding,
  title={Understanding the response to high-stakes incentives in primary education},
  author={Bach, Maximilian and Fischer, Mira},
  journal={ZEW-Centre for European Economic Research Discussion Paper},
  number={20-066},
  year={2020}
}

@article{bar2009grade,
  title={Grade information and grade inflation: The Cornell experiment},
  author={Bar, Talia and Kadiyali, Vrinda and Zussman, Asaf},
  journal={Journal of Economic Perspectives},
  volume={23},
  number={3},
  pages={93--108},
  year={2009}
}

@article{schwager2012grade,
  title={Grade inflation, social background, and labour market matching},
  author={Schwager, Robert},
  journal={Journal of Economic Behavior \& Organization},
  volume={82},
  number={1},
  pages={56--66},
  year={2012},
  publisher={Elsevier}
}

@article{betts2003impact,
  title={The impact of grading standards on student achievement, educational attainment, and entry-level earnings},
  author={Betts, Julian R and Grogger, Jeff},
  journal={Economics of Education Review},
  volume={22},
  number={4},
  pages={343--352},
  year={2003},
  publisher={Elsevier}
}

@article{figlio2004high,
  title={Do high grading standards affect student performance?},
  author={Figlio, David N and Lucas, Maurice E},
  journal={Journal of Public Economics},
  volume={88},
  number={9-10},
  pages={1815--1834},
  year={2004},
  publisher={Elsevier}
}

@article{fuchsman2020testing,
  title={Testing, teacher turnover and the distribution of teachers across grades and schools},
  author={Fuchsman, Dillon and Sass, Tim R and Zamarro, Gema},
  journal={Education Finance and Policy},
  pages={1--61},
  year={2020}
}

@article{lazear2006speeding,
  title={Speeding, terrorism, and teaching to the test},
  author={Lazear, Edward P},
  journal={The Quarterly Journal of Economics},
  volume={121},
  number={3},
  pages={1029--1061},
  year={2006},
  publisher={MIT Press}
}

@article{imberman2016does,
  title={Does the market value value-added? Evidence from housing prices after a public release of school and teacher value-added},
  author={Imberman, Scott A and Lovenheim, Michael F},
  journal={Journal of Urban Economics},
  volume={91},
  pages={104--121},
  year={2016},
  publisher={Elsevier}
}

@article{sackett2018eight,
  title={Eight myths about standardized admissions testing},
  author={Sackett, Paul R and Kuncel, Nathan R},
  journal={Measuring success: Testing, grades, and the future of college admissions},
  pages={13--39},
  year={2018},
  publisher={Johns Hopkins University Press Baltimore}
}

@article{karlson2019expectation,
  title={Expectation formation for all? Group differences in student response to signals about academic performance},
  author={Karlson, Kristian Bernt},
  journal={The Sociological Quarterly},
  volume={60},
  number={4},
  pages={716--737},
  year={2019},
  publisher={Taylor \& Francis}
}

@article{babcock2010real,
  title={Real costs of nominal grade inflation? New evidence from student course evaluations},
  author={Babcock, Philip},
  journal={Economic inquiry},
  volume={48},
  number={4},
  pages={983--996},
  year={2010},
  publisher={Wiley Online Library}
}

@article{harjunen2018best,
  title={Best education money can buy? Capitalization of school quality in Finland},
  author={Harjunen, Oskari and Kortelainen, Mika and Saarimaa, Tuukka},
  journal={CESifo Economic Studies},
  volume={64},
  number={2},
  pages={150--175},
  year={2018},
  publisher={Oxford University Press}
}

@techreport{stinebrickner2011math,
  title={Math or science? Using longitudinal expectations data to examine the process of choosing a college major},
  author={Stinebrickner, Todd R and Stinebrickner, Ralph},
  year={2011},
  institution={National Bureau of Economic Research}
}

@article{andrew2011adoption,
  title={Adoption? Adaptation? Evaluating the formation of educational expectations},
  author={Andrew, Megan and Hauser, Robert M},
  journal={Social Forces},
  volume={90},
  number={2},
  pages={497--520},
  year={2011},
  publisher={Oxford University Press}
}

@techreport{ahn2019equilibrium,
  title={Equilibrium grade inflation with implications for female interest in stem majors},
  author={Ahn, Thomas and Arcidiacono, Peter and Hopson, Amy and Thomas, James R},
  year={2019},
  institution={National Bureau of Economic Research}
}

@article{boleslavsky2015grading,
  title={Grading standards and education quality},
  author={Boleslavsky, Raphael and Cotton, Christopher},
  journal={American Economic Journal: Microeconomics},
  volume={7},
  number={2},
  pages={248--79},
  year={2015}
}

@article{fredriksson2018mismatch,
  title={Mismatch of talent: Evidence on match quality, entry wages, and job mobility},
  author={Fredriksson, Peter and Hensvik, Lena and Skans, Oskar Nordstr{\"o}m},
  journal={American Economic Review},
  volume={108},
  number={11},
  pages={3303--38},
  year={2018}
}

@article{maurin2008vive,
  title={Vive la revolution! Long-term educational returns of 1968 to the angry students},
  author={Maurin, Eric and McNally, Sandra},
  journal={Journal of Labor Economics},
  volume={26},
  number={1},
  pages={1--33},
  year={2008},
  publisher={The University of Chicago Press}
}

@article{ehlers2016honest,
  title={Honest grading, grade inflation, and reputation},
  author={Ehlers, Tim and Schwager, Robert},
  journal={CESifo Economic Studies},
  volume={62},
  number={3},
  pages={506--521},
  year={2016},
  publisher={Oxford University Press}
}

@article{zubrickas2015optimal,
  title={Optimal grading},
  author={Zubrickas, Robertas},
  journal={International Economic Review},
  volume={56},
  number={3},
  pages={751--776},
  year={2015},
  publisher={Wiley Online Library}
}

@book{smallwood1969historical,
  title={An historical study of examinations and grading systems in early American universities},
  author={Smallwood, Mary Lovett},
  year={1969},
  publisher={Harvard University Press}
}

@article{buckley2018measuring,
  title={Measuring Success: Testing, Grades, and the Future of College Admissions.},
  author={Buckley, Jack and Letukas, Lynn and Wildavsky, Ben},
  journal={Johns Hopkins University Press},
  year={2018},
  publisher={ERIC}
}

@article{grant2013grades,
  title={Grades as incentives},
  author={Grant, Darren and Green, William B},
  journal={Empirical Economics},
  volume={44},
  number={3},
  pages={1563--1592},
  year={2013},
  publisher={Springer}
}

@article{kane2002promise,
  title={The promise and pitfalls of using imprecise school accountability measures},
  author={Kane, Thomas J and Staiger, Douglas O},
  journal={Journal of Economic perspectives},
  volume={16},
  number={4},
  pages={91--114},
  year={2002}
}

@article{chan2007signaling,
  title={A signaling theory of grade inflation},
  author={Chan, William and Hao, Li and Suen, Wing},
  journal={International Economic Review},
  volume={48},
  number={3},
  pages={1065--1090},
  year={2007},
  publisher={Wiley Online Library}
}

@article{heissel2021testing,
  title={Testing, stress, and performance: How students respond physiologically to high-stakes testing},
  author={Heissel, Jennifer A and Adam, Emma K and Doleac, Jennifer L and Figlio, David N and Meer, Jonathan},
  journal={Education Finance and Policy},
  volume={16},
  number={2},
  pages={183--208},
  year={2021},
  publisher={MIT Press}
}

@article{machin2020entry,
  title={Entry through the narrow door: The costs of just failing high stakes exams},
  author={Machin, Stephen and McNally, Sandra and Ruiz-Valenzuela, Jenifer},
  journal={Journal of Public Economics},
  volume={190},
  pages={104224},
  year={2020},
  publisher={Elsevier}
}

@article{azmat2016gender,
  title={Gender differences in response to big stakes},
  author={Azmat, Ghazala and Calsamiglia, Caterina and Iriberri, Nagore},
  journal={Journal of the European Economic Association},
  volume={14},
  number={6},
  pages={1372--1400},
  year={2016},
  publisher={Oxford University Press}
}

@article{bizopoulou2022second,
  title={Do Second Chances Pay Off? Evidence from a Natural Experiment with Low-Achieving Students},
  author={Bizopoulou, Aspasia and Megalokonomou, Rigissa and Simion, Stefania},
  year={2022},
  publisher={CESifo Working Paper}
}

@article{schultz1961investment,
  title={Investment in human capital},
  author={Schultz, Theodore W},
  journal={The American economic review},
  volume={51},
  number={1},
  pages={1--17},
  year={1961},
  publisher={JSTOR}
}

@article{azmat2019you,
  title={What you don’t know… can’t hurt you? A natural field experiment on relative performance feedback in higher education},
  author={Azmat, Ghazala and Bagues, Manuel and Cabrales, Antonio and Iriberri, Nagore},
  journal={Management Science},
  volume={65},
  number={8},
  pages={3714--3736},
  year={2019},
  publisher={INFORMS}
}

@article{goodman2020take,
  title={Take two! SAT retaking and college enrollment gaps},
  author={Goodman, Joshua and Gurantz, Oded and Smith, Jonathan},
  journal={American Economic Journal: Economic Policy},
  volume={12},
  number={2},
  pages={115--58},
  year={2020}
}

@article{bond2021hungry,
  title={Hungry for Success? SNAP Timing, High-Stakes Exam Performance, and College Attendance},
  author={Bond, Timothy N and Carr, Jillian B and Packham, Analisa and Smith, Jonathan},
  year={{forthcoming}},
  journal={American Economic Journal: Economy Policy }
}

@article{diamond2016long,
  title={The long-term consequences of teacher discretion in grading of high-stakes tests},
  author={Diamond, Rebecca and Persson, Petra},
  year={2016},
  number=22207,
  journal={National Bureau of Economic Research - Working Paper}
}

@article{becker1964human,
  title={Human Capital: A Theoretical and Empirical Analysis, with Special Reference to Education},
  author={Becker, Gary S},
  year={1964}
}

@article{holmstrom1991multitask,
  title={Multitask principal-agent analyses: Incentive contracts, asset ownership, and job design},
  author={Holmstrom, Bengt and Milgrom, Paul},
  journal={JL Econ. \& Org.},
  volume={7},
  pages={24},
  year={1991},
  publisher={HeinOnline}
}

@article{heckman2002performance,
  title={The Performance of Performance Standards.},
  author={Heckman, JJ and Heinrich, C and Smith, J},
  journal={Journal of Human Resources},
  volume={37},
  number={4},
  pages={778--811},
  year={2002},
  publisher={University of Wisconsin Press}
}

@incollection{burgess2021school,
  title={School Choice and Accountability},
  author={Burgess, Simon and Greaves, Ellen},
  booktitle={Oxford Research Encyclopedia of Economics and Finance},
  year={2021}
}

@article{arrow1973higher,
  title={Higher education as a filter},
  author={Arrow, Kenneth J},
  journal={Journal of Public Economics},
  volume={2},
  number={3},
  pages={193--216},
  year={1973},
  publisher={Elsevier}
}

@article{guskey2007using,
  title={Using assessments to improve teaching and learning},
  author={Guskey, Thomas R},
  journal={Ahead of the curve: The power of assessment to transform teaching and learning},
  pages={15--29},
  year={2007},
  publisher={Solution Tree Bloomington, IN}
}

@article{beuchert2020impact,
  title={The impact of standardized test feedback in math: Exploiting a natural experiment in 3rd grade},
  author={Beuchert, Louise and Eriksen, Tine Louise Mundbjerg and Kr{\ae}gp{\o}th, Morten Visby},
  journal={Economics of Education Review},
  volume={77},
  pages={102017},
  year={2020},
  publisher={Elsevier}
}

@article{reback,
  title={Teaching to the rating: School accountability and the distribution of student achievement},
  author={Reback, Randall},
  journal={Journal of public economics},
  volume={92},
  number={5-6},
  pages={1394--1415},
  year={2008},
  publisher={Elsevier}
}

@article{borghans2015parental,
  title={Parental preferences for primary school characteristics},
  author={Borghans, Lex and Golsteyn, Bart HH and Z{\"o}litz, Ulf},
  journal={The BE Journal of Economic Analysis \& Policy},
  volume={15},
  number={1},
  pages={85--117},
  year={2015},
  publisher={De Gruyter}
}

@article{dee2019causes,
  title={The causes and consequences of test score manipulation: Evidence from the New York regents examinations},
  author={Dee, Thomas S and Dobbie, Will and Jacob, Brian A and Rockoff, Jonah},
  journal={American Economic Journal: Applied Economics},
  volume={11},
  number={3},
  pages={382--423},
  year={2019}
}

@article{burgess2015parents,
  title={What parents want: School preferences and school choice},
  author={Burgess, Simon and Greaves, Ellen and Vignoles, Anna and Wilson, Deborah},
  journal={The Economic Journal},
  volume={125},
  number={587},
  pages={1262--1289},
  year={2015},
  publisher={Wiley Online Library}
}

@article{metcalfe2019students,
  title={Students' effort and educational achievement: Using the timing of the World Cup to vary the value of leisure},
  author={Metcalfe, Robert and Burgess, Simon and Proud, Steven},
  journal={Journal of Public Economics},
  volume={172},
  pages={111--126},
  year={2019},
  publisher={Elsevier}
}

@article{graff2012impact,
  title={The impact of pollution on worker productivity},
  author={Graff Zivin, Joshua and Neidell, Matthew},
  journal={American Economic Review},
  volume={102},
  number={7},
  pages={3652--73},
  year={2012}
}

@article{sievertsen2016cognitive,
  title={Cognitive fatigue influences students’ performance on standardized tests},
  author={Sievertsen, Hans Henrik and Gino, Francesca and Piovesan, Marco},
  journal={Proceedings of the National Academy of Sciences},
  volume={113},
  number={10},
  pages={2621--2624},
  year={2016},
  publisher={National Academy Sciences}
}

@article{kunn2019indoor,
  title={Indoor air quality and cognitive performance},
  author={K{\"u}nn, Steffen and Palacios, Juan and Pestel, Nico},
  year={{forthcoming}},
  journal={Management Science}
}

@article{hyman2017act,
  title={ACT for all: The effect of mandatory college entrance exams on postsecondary attainment and choice},
  author={Hyman, Joshua},
  journal={Education Finance and Policy},
  volume={12},
  number={3},
  pages={281--311},
  year={2017},
  publisher={MIT Press One Rogers Street, Cambridge, MA 02142-1209, USA journals-info~…}
}

@article{hurwitz2015maine,
  title={The Maine question: How is 4-year college enrollment affected by mandatory college entrance exams?},
  author={Hurwitz, Michael and Smith, Jonathan and Niu, Sunny and Howell, Jessica},
  journal={Educational Evaluation and Policy Analysis},
  volume={37},
  number={1},
  pages={138--159},
  year={2015},
  publisher={SAGE Publications Sage CA: Los Angeles, CA}
}

@article{murphy2020minority,
  title={Minority Report: the impact of predicted grades on university admissions of disadvantaged groups},
  author={Murphy, Richard and Wyness, Gill},
  journal={Education Economics},
  volume={28},
  number={4},
  pages={333--350},
  year={2020},
  publisher={Taylor \& Francis}
}

@article{borghans2016grades,
  title={What grades and achievement tests measure},
  author={Borghans, Lex and Golsteyn, Bart HH and Heckman, James J and Humphries, John Eric},
  journal={Proceedings of the National Academy of Sciences},
  volume={113},
  number={47},
  pages={13354--13359},
  year={2016},
  publisher={National Acad Sciences}
}

@article{denning2018class,
  title={Class rank and long-run outcomes},
  author={Denning, Jeffrey T and Murphy, Richard and Weinhardt, Felix},
  journal={The Review of Economics and Statistics},
  pages={1--45},
  year={2018}
}

@article{elsner2017big,
  title={A big fish in a small pond: Ability rank and human capital investment},
  author={Elsner, Benjamin and Isphording, Ingo E},
  journal={Journal of Labor Economics},
  volume={35},
  number={3},
  pages={787--828},
  year={2017},
  publisher={University of Chicago Press Chicago, IL}
}

@article{hvidman2021high,
  title={High-Stakes Grades and Student Behavior},
  author={Hvidman, Ulrik and Sievertsen, Hans Henrik},
  journal={Journal of Human Resources},
  volume={56},
  number={3},
  pages={821--849},
  year={2021},
  publisher={University of Wisconsin Press}
}

@article{murphy2020top,
  title={Top of the class: The importance of ordinal rank},
  author={Murphy, Richard and Weinhardt, Felix},
  journal={The Review of Economic Studies},
  volume={87},
  number={6},
  pages={2777--2826},
  year={2020},
  publisher={Oxford University Press}
}

@article{andersen2020learning,
  title={Learning from performance information},
  author={Andersen, Simon Calmar and Nielsen, Helena Skyt},
  journal={Journal of Public Administration Research and Theory},
  volume={30},
  number={3},
  pages={415--431},
  year={2020},
  publisher={Oxford University Press US}
}

@article{stinebrickner2012learning,
  title={Learning about academic ability and the college dropout decision},
  author={Stinebrickner, Todd and Stinebrickner, Ralph},
  journal={Journal of Labor Economics},
  volume={30},
  number={4},
  pages={707--748},
  year={2012},
  publisher={University of Chicago Press Chicago, IL}
}

@article{burgess2013test,
  title={Test scores, subjective assessment, and stereotyping of ethnic minorities},
  author={Burgess, Simon and Greaves, Ellen},
  journal={Journal of Labor Economics},
  volume={31},
  number={3},
  pages={535--576},
  year={2013},
  publisher={University of Chicago Press Chicago, IL}
}

@article{terrier2020boys,
  title={Boys lag behind: How teachers’ gender biases affect student achievement},
  author={Terrier, Camille},
  journal={Economics of Education Review},
  volume={77},
  pages={101981},
  year={2020},
  publisher={Elsevier}
}

@article{rangvid2015systematic,
  title={Systematic differences across evaluation schemes and educational choice},
  author={Rangvid, Beatrice Schindler},
  journal={Economics of Education Review},
  volume={48},
  pages={41--55},
  year={2015},
  publisher={Elsevier}
}

@article{nordin2019impact,
  title={The impact of grade inflation on higher education enrolment and earnings},
  author={Nordin, Martin and Heckley, Gawain and Gerdtham, Ulf},
  journal={Economics of Education Review},
  volume={73},
  pages={101936},
  year={2019},
  publisher={Elsevier}
}

@article{sabot1991grade,
  title={Grade inflation and course choice},
  author={Sabot, Richard and Wakeman-Linn, John},
  journal={Journal of Economic Perspectives},
  volume={5},
  number={1},
  pages={159--170},
  year={1991}
}

@article{falch2013educational,
  title={Educational evaluation schemes and gender gaps in student achievement},
  author={Falch, Torberg and Naper, Linn Ren{\'e}e},
  journal={Economics of Education Review},
  volume={36},
  pages={12--25},
  year={2013},
  publisher={Elsevier}
}

@article{jackson2018test,
  title={What do test scores miss? The importance of teacher effects on non--test score outcomes},
  author={Jackson, C Kirabo},
  journal={Journal of Political Economy},
  volume={126},
  number={5},
  pages={2072--2107},
  year={2018},
  publisher={University of Chicago Press Chicago, IL}
}

@article{burgess2022importance,
  title={The importance of external assessments: High school math and gender gaps in STEM degrees},
  author={Burgess, Simon and Hauberg, Daniel Sloth and Rangvid, Beatrice Schindler and Sievertsen, Hans Henrik},
  journal={Economics of Education Review},
  volume={88},
  pages={102267},
  year={2022},
  publisher={Elsevier}
}

@article{LAVY20082083,
title = {Do gender stereotypes reduce girls' or boys' human capital outcomes? Evidence from a natural experiment},
journal = {Journal of Public Economics},
volume = {92},
number = {10},
pages = {2083-2105},
year = {2008},
issn = {0047-2727},
doi = {https://doi.org/10.1016/j.jpubeco.2008.02.009},
url = {https://www.sciencedirect.com/science/article/pii/S0047272708000418},
author = {Victor Lavy}
}

@techreport{ofs2022,
  title={Analysis of degree classifications over time - Changes in graduate attainment from 2010-11 to 2020-21},
  author={{OFS}},
  year={2022},
  institution={Office for Students}
}

@techreport{lavy2019persistency,
  title={Persistency in teachers’ grading bias and effects on longer-term outcomes: University admissions exams and choice of field of study},
  author={Lavy, Victor and Megalokonomou, Rigissa},
  year={2019},
  institution={National Bureau of Economic Research}
}

@article{elsner2021achievement,
  title={Achievement rank affects performance and major choices in college},
  author={Elsner, Benjamin and Isphording, Ingo E and Z{\"o}litz, Ulf},
  journal={The Economic Journal},
  volume={131},
  number={640},
  pages={3182--3206},
  year={2021},
  publisher={Oxford University Press}
}

@article{elsner2018rank,
  title={Rank, sex, drugs, and crime},
  author={Elsner, Benjamin and Isphording, Ingo E},
  journal={Journal of Human Resources},
  volume={53},
  number={2},
  pages={356--381},
  year={2018},
  publisher={University of Wisconsin Press}
}

@article{hanna2012discrimination,
  title={Discrimination in grading},
  author={Hanna, Rema N and Linden, Leigh L},
  journal={American Economic Journal: Economic Policy},
  volume={4},
  number={4},
  pages={146--68},
  year={2012}
}

@article{dee2005teacher,
  title={A teacher like me: Does race, ethnicity, or gender matter?},
  author={Dee, Thomas S},
  journal={American Economic Review},
  volume={95},
  number={2},
  pages={158--165},
  year={2005}
}

@article{lindahl2016teacher,
  title={Are teacher assessments biased?--evidence from Sweden},
  author={Lindahl, Erica},
  journal={Education economics},
  volume={24},
  number={2},
  pages={224--238},
  year={2016},
  publisher={Taylor \& Francis}
}

@article{gneezy2003performance,
  title={Performance in competitive environments: Gender differences},
  author={Gneezy, Uri and Niederle, Muriel and Rustichini, Aldo},
  journal={Quarterly Journal of Economics},
  volume={118},
  number={3},
  pages={1049--1074},
  year={2003},
  publisher={MIT Press}
}

@article{goldin2000orchestrating,
  title={Orchestrating impartiality: The impact of" blind" auditions on female musicians},
  author={Goldin, Claudia and Rouse, Cecilia},
  journal={American economic review},
  volume={90},
  number={4},
  pages={715--741},
  year={2000}
}

@article{phelps1972statistical,
  title={The statistical theory of racism and sexism},
  author={Phelps, Edmund S},
  journal={American Economic Review},
  volume={62},
  number={4},
  pages={659--661},
  year={1972},
  publisher={JSTOR}
}

@incollection{almlund2011personality,
  title={Personality psychology and economics},
  author={Almlund, Mathilde and Duckworth, Angela Lee and Heckman, James and Kautz, Tim},
  booktitle={Handbook of the Economics of Education},
  volume={4},
  pages={1--181},
  year={2011},
  publisher={Elsevier}
}

@article{schleicher2014oecd,
  title={Letter by academics - OECD and Pisa tests are damaging education worldwide-academics},
  author={{The Guardian}},
  journal={Guardian Education},
  year={2014}
}

@book{wb2018,
  title={World Development Report -Learning to realize education's promise},
  author={{The World Bank}},
  year={2018},
  publisher={World Bank Group}
}

\end{document}